\documentclass[10pt,conference]{IEEEtran}
\IEEEoverridecommandlockouts

\usepackage{times}
\usepackage{latexsym}
\usepackage{balance}

\usepackage[T1]{fontenc}
\usepackage{microtype}
\usepackage{fancyhdr}
\usepackage{inconsolata}

\usepackage{amsmath,amssymb,amsfonts}
\usepackage{algorithmic}
\usepackage{graphicx}
\usepackage{textcomp}
\usepackage{xcolor}
\def\BibTeX{{\rm B\kern-.05em{\sc i\kern-.025em b}\kern-.08em
    T\kern-.1667em\lower.7ex\hbox{E}\kern-.125emX}}

\title{SparseCoder: Identifier-Aware Sparse Transformer for File-Level Code Summarization}

\usepackage{enumitem}
\usepackage{makecell}
\usepackage{xspace}
\usepackage{amsmath,amssymb,amsfonts}
\usepackage{graphicx}
\usepackage{textcomp}
\usepackage{subfigure}
\usepackage{tcolorbox}
\usepackage{booktabs}
\usepackage{tabularx}
\usepackage{lipsum}
\usepackage{multirow}
\usepackage{xcolor}
\newcommand{\newrevised}[1]{\textcolor{black}{#1}}
\newcommand{\revised}[1]{\textcolor{black}{#1}}
\newcommand{\myauthornote}[3]{}

\newcommand{\later}[1]{}

\usepackage{CJKutf8}

\usepackage{listings}
\usepackage{xcolor}
\definecolor{light-gray}{gray}{0.99}
\usepackage{fancybox}


\newcommand{\saveSpaceSec}{\vspace{0pt}}
\newcommand{\incSpace}{\vspace{3pt}}
\newcommand{\dataset}{FILE-CS\xspace} 
\newcommand{\model}{SparseCoder\xspace} 

\usepackage{tikz}
\newcommand*{\circled}[1]{\lower.7ex\hbox{\tikz\draw (0pt, 0pt)%
    circle (.5em) node {\makebox[1em][c]{\small #1}};}}

\definecolor{light-gray}{gray}{0.95}    

\author{
    \IEEEauthorblockN{
        Yanlin Wang\IEEEauthorrefmark{2}, 
        Yanxian Huang\IEEEauthorrefmark{2}, 
        Daya Guo\IEEEauthorrefmark{4}, 
        Hongyu Zhang\IEEEauthorrefmark{1}\IEEEauthorrefmark{3} \thanks{* Corresponding author.}
        and Zibin Zheng\IEEEauthorrefmark{2}} 
	\IEEEauthorblockA{\IEEEauthorrefmark{2}School of Software Engineering, Sun Yat-sen University, China}
	\IEEEauthorblockA{\IEEEauthorrefmark{4}School of Computer Science and Technology, Sun Yat-sen University, China}
    \IEEEauthorblockA{\IEEEauthorrefmark{3}School of Big Data and Software Engineering, Chongqing University, China}
     
	\IEEEauthorblockA{\{wangylin36, zhzibin\}@mail.sysu.edu.cn, \{huangyx353, guody5\}@mail2.sysu.edu.cn, hyzhang@cqu.edu.cn} 
	
}

\usepackage{biblatex}
\begin{document}
\maketitle
\pagestyle{plain}
\begin{abstract}
Code summarization aims to generate natural language descriptions of source code, facilitating programmers to understand and maintain it rapidly. While previous code summarization efforts have predominantly focused on method-level, this paper studies file-level code summarization, which can assist programmers in understanding and maintaining large source code projects. Unlike method-level code summarization, file-level code summarization typically involves long source code within a single file, which makes it challenging for Transformer-based models to understand the code semantics for the maximum input length of these models is difficult to set to a large number that can handle long code input well, due to the quadratic scaling of computational complexity with the input sequence length.
To address this challenge, we propose SparseCoder, an identifier-aware sparse transformer for effectively handling long code sequences. Specifically, the \model employs a sliding window mechanism for self-attention to model short-term dependencies and leverages the structure message of code to capture long-term dependencies among source code identifiers by introducing two types of sparse attention patterns named global and identifier attention.
To evaluate the performance of SparseCoder, we construct a new dataset \dataset for file-level code summarization in Python. Experimental results show that our \model model achieves state-of-the-art performance compared with other pre-trained models, including full self-attention and sparse models. Additionally, our model has low memory overhead and achieves comparable performance with models using full self-attention mechanism. Furthermore, we verify the generality of SparseCoder on other code understanding tasks, i.e., code clone detection and code search, and results show that our model outperforms baseline models in both tasks, demonstrating that our model can generate better code representations for various downstream tasks.
Our source code and experimental data are anonymously available at: \url{https://github.com/DeepSoftwareAnalytics/SparseCoder}
\end{abstract}


\section{Introduction}
Source code summarization aims to generate a natural language summary from a piece of code, which is a significant task in software engineering that can assist developers in comprehending and maintaining codebases. 
\newrevised{Additionally, it contributes to other tasks like code clone detection and code search, enhancing overall efficiency in software development. Specifically, the generated summaries can help identify code clones more easily or search the code better.}
Different from previous works~\cite{AhmadCRC20,haije2016automatic,HuLXLJ18,hu2019deep,HuLXLLJ18,IyerKCZ16,LeClairJM19,WanZYXY0Y18,YuHWF020} that mainly focus on method-level code summarization, i.e., generating natural language summaries for individual methods,
we study the task of file-level code summarization in this paper. The goal of file-level code summarization is to generate a high-level summary for an entire code file rather than individual methods. This can help developers save a significant amount of time and increase efficiency in software development, as file-level summarization makes it easier for them to navigate and understand the codebase, and make changes or updates as needed. Without file-level code summarization, a developer would have to manually open and read through each file to understand its contents and how it relates to the large project. 
Compared to method-level summarization, file-level code summarization typically involves inputting longer source code within a single file, making it challenging for pre-trained code language models~\cite{Ahmad2021UnifiedPF,feng2020codebert,guo2021graphcodebert,guo-etal-2022-unixcoder,Wang2021CodeT5IU} that achieves state-of-the-art performance on various code-related downstream tasks.
It is difficult for the models to handle the task effectively with the long input sequence due to the quadratic scaling of the self-attention mechanism's computational complexity with the input sequence length. One approach to handle long input sequences is to truncate them to a maximum number of tokens, such as 512 or 1024. However, this method may not be sufficient as the average length of a Python language source code file on GitHub is 2,090 tokens after using the code tokenizer released by existing study\cite{guo-etal-2022-unixcoder}, and 41\% of the files have a length longer than 1,024 tokens. Truncating the sequence in this manner can result in important information being omitted.

To address this limitation, we propose \textbf{\model}, an identifier-aware sparse Transformer designed to efficiently model long code sequences. \model is built on the Transformer architecture but employs local attention to model short-term dependencies and leverages the structure of code to capture long-term dependencies among identifiers in source code. 
For \emph{local attention}, we adopt the sliding window mechanism\cite{beltagy2020longformer} to model short-term dependencies, which allows each token to attend only to its neighboring tokens. This local attention mechanism is sufficient for most tokens in code, such as keywords and literals, which mainly depend on local context. Furthermore, this local attention mechanism ensures that the computational complexity of the model scales linearly with the input sequence length and enables \model to handle longer sequences. 
To effectively model long-range dependencies among identifiers with long scope, such as libraries and global variables, we introduce two types of sparse attention patterns: \emph{global attention} and \emph{identifier attention}. 
Global attention identifies identifiers with a global scope through static analysis and considers them as global tokens, allowing all tokens to attend to them and vice versa. These global identifiers are typically libraries, classes, functions, and global variables. 
For other identifiers that may have potential long-range dependencies, identifier attention applies self-attention operations among them to capture long-term dependencies. 
To enhance the flexibility of \model in modeling different attention types, we introduce a low-rank adaptation approach for each linear projection within the self-attention layer.

To evaluate the performance of \model, we construct a new dataset called  \textbf{\dataset} for file-level code summarization, including 98,236 <code file, summary> pairs in Python. Experimental results show that our model achieves state-of-the-art performance compared with other pre-trained models, demonstrating the effectiveness of the combination of various types of attention mechanisms, and the results of the ablation experiments show the effectiveness of the proposed global and identifier attention. Further analysis reveals that our model has low memory overhead and performs comparably to models using full self-attention and other sparse models. Additionally, our \model model can also generalize to other code understanding tasks like code clone detection and code search.

In summary, this paper makes the following contributions: 
\begin{itemize}
\item We propose \model, an identifier-aware sparse Transformer model that can efficiently handle long code sequences and obtain better code representation.  
\item We create \dataset, the first dataset for the task of file-level code summarization, and will make it available to the research community. 
\item Experimental results show that \model achieves significant improvement in various downstream tasks, confirming its effectiveness.
\end{itemize}

\section{Related Work}
\label{sec:related_work}
\subsection{Code summarization}
Over the years, researchers have proposed numerous approaches to code summarization, which can generally be classified into three categories: rule-based, information retrieval (IR)-based, and learning-based approaches.

\subsubsection{Rule-based and IR-based approaches}
Early work in this area largely employed rule-based and IR-based approaches~\cite{EddyRKC13,HaiducAM10ICSE,HaiducAMM10WCRE, mcburney2014automatic,Movshovitz-AttiasC13,RodegheroMMBD14,SridharaHMPV10}.  
Sridhara et al.~\cite{SridharaHMPV10} design heuristics to choose statements from Java methods and use the Software Word Usage Model (SWUM) to identify keywords from those statements and create summaries through  manually-defined templates. \newrevised{Eddy et al.~\cite{EddyRKC13} propose a new topic modeling based approach to source code summarization and Movshovitz-Attias et al.~\cite{Movshovitz-AttiasC13} predict comments from
Java source files of open source projects, using topic models and n-grams. Rodeghero et al.~\cite{RodegheroMMBD14} present an eye-tracking study of ten professional Java programmers reading Java methods and writing English summaries of these methods and then apply the findings to build a novel summarization tool. } 
Paul et al.~\cite{mcburney2014automatic} use PageRank to choose important methods that are invoked by the target method and apply SWUM to extract program keywords to generate summaries. 
Haiduc et al.~\cite{HaiducAM10ICSE,HaiducAMM10WCRE} treat each function of source code as a document and index on such a corpus by LSI or VSM, and select the most similar terms based on the cosine distances between documents as the summary. 
Rule-based approaches for code summarization suffer from the limitation of being highly dependent on the specific programming language to be summarized, making them difficult to generalize. IR-based approaches, relying on text matching, 
may exhibit limitations in capturing the code semantics.

\subsubsection{Learning-based approaches}
Learning-based approaches have also gained widespread use in code summarization~\cite{AhmadCRC20,haije2016automatic,HuLXLJ18,hu2019deep,HuLXLLJ18,IyerKCZ16,LeClairJM19,WanZYXY0Y18,YuHWF020}. 
Iyer et al.~\cite{IyerKCZ16} propose the first neural approach for code summarization, using a classical encoder-decoder framework with an attention mechanism to encode code into context vectors and generate summaries in the decoder.  \newrevised{Since then, a series of work~\cite{haije2016automatic,HuLXLJ18,hu2019deep} has been done on automatically generating summaries of code using a deep neural network. Hu et al.~\cite{HuLXLLJ18}  propose a novel approach using API knowledge learned in a different but related task to code summarization. LeClair et al.~\cite{LeClairJM19} present a neural model that combines words from code with code structure from an AST.} Ahmad et al.~\cite{AhmadCRC20} model code using Transformer to capture the long-range dependencies, while Wan et al.~\cite{WanZYXY0Y18} employ hybrid code representations and deep reinforcement learning to encode both the sequential and structural content of code and use a hybrid attention layer to obtain an integrated representation. 
To more accurately capture the structural information of the source code, Gong et al.~\cite{Gong2022SourceCS} propose to capture the structural relative positions between tokens for better learning code semantics. Gao et al.~\cite{Gao2022M2TSMM} use a multi-scale AST feature extraction method, which can extract the structures of ASTs more completely and accurately at multiple local and global levels. Guo et al.~\cite{Guo2022ModelingHS} model the hierarchical syntax structure of code by introducing a triplet position and propose a pointer-generator network for better summary generation. 
Son et al.~\cite{Son2022BoostingCS} propose a PDG Boosting Module (PBM) that encodes a program dependence graph into a graph embedding to improve the performance of the existing models. Gao et al.~\cite{Gao2022GTSimNetIC} propose a code semantic modeling method based on a local application programming interface (API) dependency graph (Local-ADG), which exhibits an excellent ability to mask irrelevant semantics outside the current code snippet. Wang et al.~\cite{Wang2022GypSumLH} propose a new type of semantic edges for connecting AST nodes to turn AST into semantic graphs and use graph attention neural networks to learn from the constructed semantic graphs to capture the key elements in the graphs more effectively. 

With the great success of
pre-training in code intelligence~\cite{Ahmad2021UnifiedPF,feng2020codebert,guo2021graphcodebert,guo-etal-2022-unixcoder,Wang2021CodeT5IU}, pre-trained code models have been used to improve the performance of code summarization. CodeBERT~\cite{feng2020codebert} and CuBERT~\cite{kanade2020learning} are based on  BERT~\cite{devlin2018bert} and trained on natural languages and code using masked language modeling. CodeT5~\cite{Wang2021CodeT5IU} is a 
unified model with a bimodal dual generation task to better align NL-PL. GraphCodeBERT~\cite{guo2021graphcodebert} and UniXcoder~\cite{guo-etal-2022-unixcoder} incorporate code-specific properties such as abstract syntax trees or data flows into model pre-training.  
However, previous works mainly apply pre-trained models to method-level code summarization, where the input code length is usually short. In contrast, this paper studies the task of file-level code summarization, which requires the model to handle longer code efficiently.

\subsection{Sparse Transformers}
Recently, techniques for handling long sequences in natural language processing (NLP) have garnered much attention~\cite{beltagy2020longformer,child2019generating,dai2019transformer,qiu2019blockwise,wu2021hi,zaheer2020big}. For example, BlockBERT~\cite{qiu2019blockwise} divides the attention matrix into $k$ blocks and defines attention within each block, reducing computational and memory cost. Sparse Transformer~\cite{child2019generating} and Longformer~\cite{beltagy2020longformer} use sliding windows and global tokens to integrate local and global information from input sequences. BigBird~\cite{zaheer2020big} builds upon Sparse Transformer by incorporating random attention, and Hi-Transformer~\cite{wu2021hi} models long documents in a hierarchical manner through the combination of sentence Transformers and document transformers. \revised{LongT5~\cite{guo-etal-2022-longt5} proposes a new attention mechanism named TGlobal (Transient Global Attention), which enables input tokens to interact with each other at a longer range compared to Local Attention's local radius. TGlobal divides the input sequence into blocks of k tokens and computes the global token for each block by summing and normalizing the embeddings of all tokens in the block. LSG Attention~\cite{condevaux2023lsg} introduces the LSG architecture, which leverages Local, Sparse, and Global attention mechanisms. The Local attention mechanism in LSG Attention is the same as BigBird. The Sparse attention mechanism extends the local context by selecting an additional set of tokens based on specific rules (selecting
based on specific metrics directly or using some computation such as a pooling method). In addition, the Global attention mechanism enables all tokens across the sequence to attend to each other, prompting bidirectional interactions among all tokens.}

While these approaches have been effective in processing long \emph{natural language} sequences, they are not specifically designed for \emph{code}-related tasks. The handling of long \emph{code} sequences has received less attention. Liu et al. \cite{liu2022understanding} propose SASA, a structure-aware sparse attention mechanism for long code understanding tasks including code summarization, which uses top-k sparse attention and AST-based structure-aware attention to capture important attention relations in code with a lower computational cost. However, the top-k attention mechanism used in SASA is based on the assumption of a fixed list of variables, which can be problematic because variable names are not the invariants for code semantics. That is, performing a variable renaming operation on a piece of code does not change its semantics. Therefore, it can be difficult to accurately identify the most important parts of the code using the top-k attention mechanism. In contrast, our work uses an identifier attention mechanism, which allows the model to dynamically learn the crucial parts of the code without the fixed variable name assumption. 

\begin{figure*}[t]
    \centering
    \includegraphics[width=0.92\linewidth]{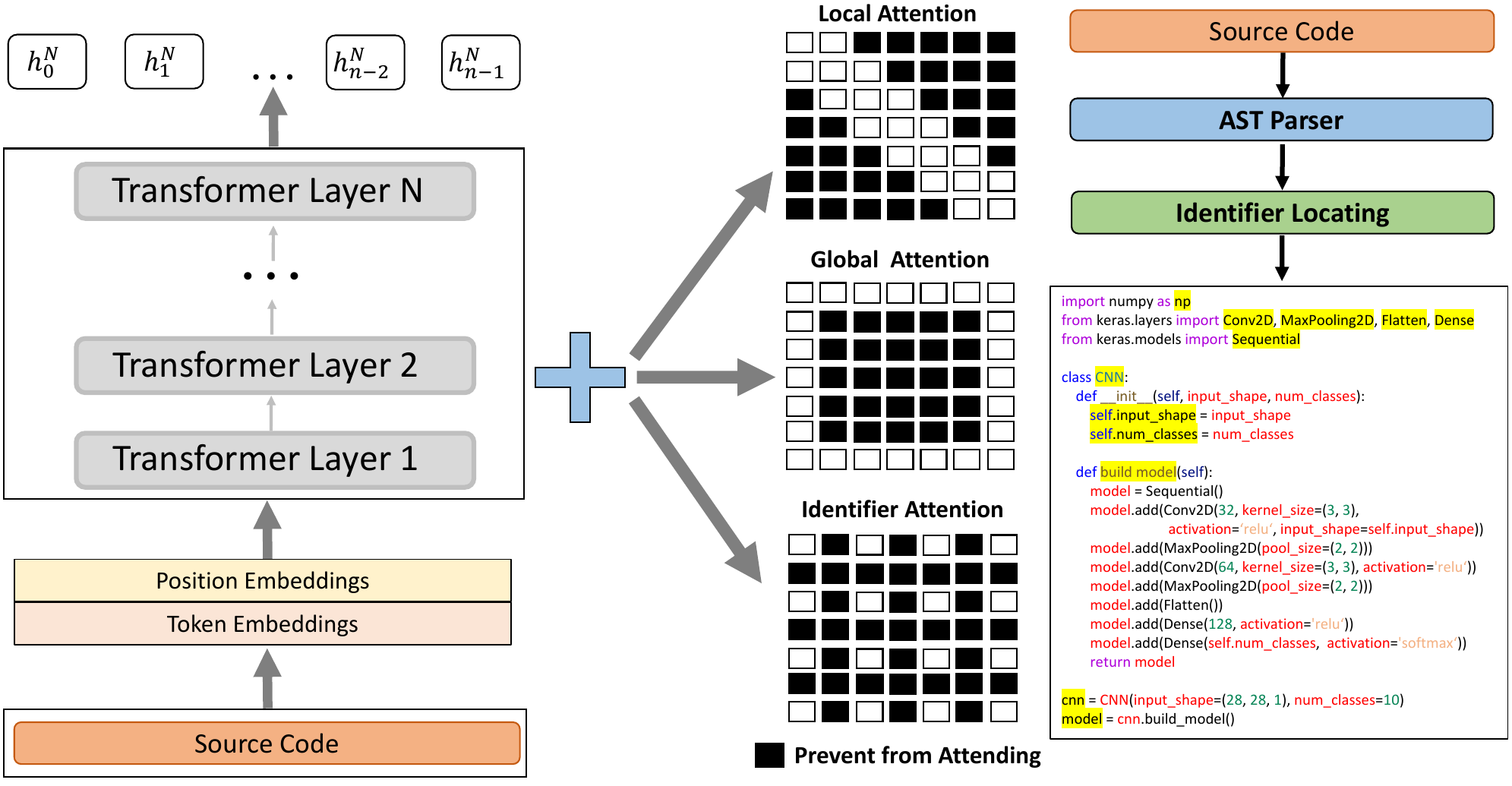}
    \caption{The overview of \model. Global and local identifiers are marked in yellow and red, respectively.
}
\vspace{-0.2in}
\saveSpaceSec
\label{fig:model}
\end{figure*}

\section{SparseCoder}
The original Transformer model has a computational and memory complexity of $O(n^2)$ due to its use of the self-attention mechanism, where $n$ is the length of the input sequence.
However, a file usually contains thousands of tokens, and the quadratic complexity makes it difficult to utilize long code sequences effectively. To address this challenge, we introduce \model, an identifier-aware sparse Transformer model for efficiently modeling longer code sequences. \model leverages the structure message of code to improve the self-attention mechanism by introducing three types of sparse attention patterns: local attention, global attention, and identifier attention. In addition, the \model improves efficiency by using the low-rank adaptation method (LoRA) to reduce parameters. In this section, we will introduce the model architecture of \model, three types of sparse attention patterns, and the LoRA method adopted in \model.

\subsection{Model Architecture}
Figure~\ref{fig:model} shows the model architecture of \model. The model is based on the original Transformer~\cite{vaswani2017attention} and consists of N transformer layers, which are applied to source code to obtain its hidden states $H^N=\{h_0^N,h_1^N,...,h_{n-1}^N\}$, where $n$ is the length of the input code sequence. Each transformer layer contains an architecturally identical transformer that applies a multi-headed self-attention operation~\cite{vaswani2017attention} followed by a feed forward layer.
The output of the multi-headed self-attention for the $l$-th transformer layer is computed as follows:
\begin{equation}\label{eq:projection}
Q=H^{l-1}W^{Q},K=H^{l-1}W^{K},V=H^{l-1}W^{V}
\end{equation}
\begin{equation}\label{eq:score}
head=\rm{softmax}(\frac{QK^T}{\sqrt{d_k}}+M)V
\end{equation}
where the output of the multi-headed self-attention in the $l$-th transformer layer is computed using the previous layer's output $H^{l-1}\in\mathbb{R}^{n \times d_h}$, which is mapped linearly to queries, keys, and values using model parameters $W^{Q}$, $W^{K}$, $W^{V}\in\mathbb{R}^{d_h \times d_k}$, respectively. $d_k$ is the dimension of a head, and $M\in\mathbb{R}^{n \times n}$ is a mask matrix that controls which context a token can attend to when computing its contextual representation, as shown in the middle of Figure \ref{fig:model}. If the $i$-th token is allowed to attend to the $j$-th token, then set $M_{ij}$ to 0, otherwise $-\infty$. When $M_{ij}$ is set to $-\infty$, the attention score from the $i$-th to $j$-th token will be zero after using the softmax function. This can help reduce computation by avoiding the calculation of the attention score between these tokens.

\begin{figure}[ht]
    \centering
    \includegraphics[width=0.70\linewidth]{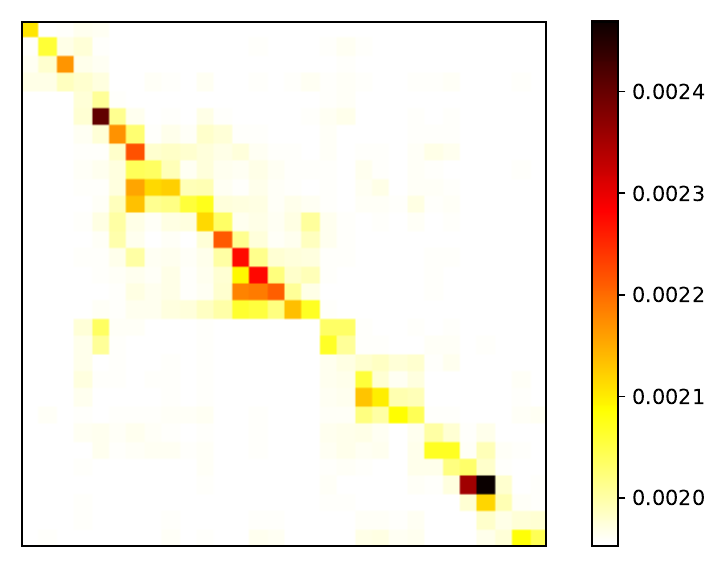}
    \caption{A code example of the attention matrix map.}
    \label{fig:att_map}
    \vspace{-0.2in}
    \saveSpaceSec
\end{figure}

For source code, we observe that the attention matrix is very sparse, and the attention scores are mainly concentrated on the diagonal of the matrix and some specific tokens, as shown in Figure \ref{fig:att_map}. Based on these observations and the structural characteristics of code, we introduce three attention patterns to reduce computational complexity: local attention, global attention, and identifier attention. 

\label{sec:local}
\subsection{Local Attention}
\newrevised{Figure 2 is a heat map obtained by calculating the attention matrix score between each token of the code sample. Darker color means higher score.}
The result of Figure \ref{fig:att_map} shows that the attention between code tokens is usually sparse and primarily concentrated along the diagonal, which is consistent with the inherent property of code where dependent tokens often appear close to each other. For example, most keywords like \textit{``import''} and \textit{``=''} in source code typically depend only on local context rather than long-range dependencies. In addition, local variables like \textit{``cnn''} in Figure \ref{fig:model} are defined and immediately used, which require only local attention. Following previous works~\cite{beltagy2020longformer,zaheer2020big}, we employ a fixed-size sliding window attention surrounding each token as our local attention. Given a fixed window size $w$, each token can attend to $\lfloor \frac{w}{2} \rfloor$ tokens on both sides for each transformer layer, and the mask matrix $M^{local}$ of local attention  is calculated as follows:
\begin{equation}
    M^{local}_{ij} = 
    \begin{cases}
        0 & \text{if}\ |i-j| \leq  \lfloor \frac{w}{2} \rfloor \\
        -\infty & \text{otherwise}
    \end{cases}
\end{equation}

The sliding window attention pattern reduces the complexity of the self-attention mechanism by limiting the receptive field size of each token to a small window of size $w$ at each layer. Note that we only need to calculate the attention score from $i$-th token to $j$-th token when $M^{local}_{ij}$ is zero. Thus, the computation complexity of this local attention pattern is $O(n \times w)$ with $w\ll n$, which scales linearly with the input sequence length $n$. After applying N transformer layers of such local attention, the receptive field size at the top layer is $N\times w$. 

\subsection{Global Attention}
In source code, identifiers with global scope, such as libraries, classes, functions, and global variables, can be called from any location within a file. This makes local attention insufficient for understanding code segments that call these global identifiers from a distance beyond the window size.  Take Figure \ref{fig:model} as an example. The code creates a variable \textit{``cnn''} by instantiating an object from the \textit{``CNN''} class on the second to last line. However, the definition of the class is located at the beginning of the file, while the creation of the variable at the end of the file. They are far apart, and the model may struggle to understand the semantics of this variable because local attention cannot effectively capture the long-range dependencies between the definition of the class and the creation of the variable. Therefore, we introduce a global attention mechanism that leverages the structure message of code to help capture the long-range dependencies.

Specifically, we use a standard compiler tool\footnote{\url{https://github.com/tree-sitter/tree-sitter}} to parse the code into an AST and find identifiers with global scope through static analysis. These global identifiers usually are library, class, function, and global variable names. An example is shown in Figure \ref{fig:model}, where global identifiers are marked in yellow. Since some global identifiers have multiple sub-tokens after tokenization, this will cause additional computation in global attention. Therefore, we only consider the first sub-token of each global identifier as the global token and denote the positions of these global tokens as $G$. 
We follow previous work~\cite{beltagy2020longformer} to make this attention symmetric, allowing global tokens to attend to all tokens and all tokens to attend to them. This is particularly important for class and function names, as they typically have longer bodies and require a longer receptive field size. Finally, the mask matrix $M^{global}$ of global attention is calculated as follows:
\begin{equation}
    M^{global}_{ij} = 
    \begin{cases}
        0 & \text{if}\ i\ or\ j\ \in G \\
        -\infty & \text{otherwise}
    \end{cases}
\end{equation}

Since we only need to calculate the attention score between the $i$-th and $j$-th tokens when one of them is a global identifier, the computation complexity of this global attention is $O(n \times |G|)$. Moreover, only a few tokens are global identifiers within a file, and the complexity scales linearly with the input sequence length $n$.

\subsection{Identifier Attention}
Generally, code contains three types of tokens: keywords, identifiers, and literals. Keywords like \textit{``=''} and literals like \textit{``relu''} typically only have short-range dependencies and can be effectively handled by local attention. However, identifiers such as parameters \textit{``input\_shape''} in Figure \ref{fig:model} may have longer-range dependencies that extend beyond the window size, and as they are not global tokens, they cannot be effectively modeled by the global attention. Therefore, we propose identifier attention that enables the model to consider long-range dependencies among identifiers. Specifically, we use AST to identify all identifiers in the code and allow them to attend to each other in a similar manner as full self-attention. As with global attention, we only consider the first sub-token of each identifier after tokenization, and denote the positions of these identifiers as $I$. The mask matrix $M^{identifier}$ of identifier attention is calculated as follows:
\begin{equation}
    M^{identifier}_{ij} = 
    \begin{cases}
        0 & \text{if}\ i\ and\ j\ \in I-G \\
        -\infty & \text{otherwise}
    \end{cases}
\end{equation}
Note that global identifiers in $I$ have already been attended to each other in the global attention, thus they do not need to participate in identifier attention. 

Since we only need to calculate the attention scores among identifiers, the computation complexity of identifier attention is $O(|I-G|^2)$. Based on our statistics, the average size of $|I-G|$ is approximately 16\% of $n$ after tokenization. If we set the maximum size of $|I-G|$ to be 25\% of $n$, we will be able to cover all identifiers in approximately 95\% of the files and the computation complexity of this identifier attention will become $O(\frac{1}{16}n^2)$. Compared to full attention, which considers the entire input sequence, identifier attention is 16 times more efficient in computation.

By introducing the three attention patterns described above, the \model can efficiently encode long code sequences. The attention mask matrix $M$ in Equation \ref{eq:score} becomes:
\begin{equation}
    M = max(M^{local}_{ij}, M^{global}_{ij}, M^{identifier}_{ij})
\end{equation}
\noindent where $max$ is element-wise maximum function.

\subsection{Low-Rank Adaptation Method for Global Attention}\label{sec:lora}

In Equation \ref{eq:projection}, LongFormer~\cite{beltagy2020longformer} employs two sets of projections \{$W^Q$ , $W^K$ , $W^V$\} to calculate attention scores for local attention, and \{$W_g^Q$ , $W_g^K$ , $W_g^V$\} to compute attention scores for global attention. The additional projections provide flexibility in modeling different attention types, which has been demonstrated to be crucial for achieving optimal performance on downstream tasks. However, this method introduces approximately 16\%  parameters of the original model. Alternatively,  we employ a low-rank adaptation method \cite{hulora} for global attention, which achieves comparable performance with much fewer parameters, approximately 0.5\% of the original model. Taking $W_g^Q$ for an example, we represent the linear projection $W_g^Q$ for global attention as the sum of $W^Q$ for other attentions and a low-rank decomposition:
\begin{equation}
W_g^Q = W^Q + \Delta W_g^Q \approx W^Q + B^QA^Q
\end{equation}
where $W_g^Q$, $W^Q$, $\Delta W_g^Q \in \mathbb{R}^{d_h\times d_k}$, $B^Q \in \mathbb{R}^{d_h\times r}$ and $A^Q \in \mathbb{R}^{r\times d_k}$ are model parameters with rank $r \ll \min(d_h,d_k)$. We initialize $A^Q$ randomly using a Gaussian distribution and $B^Q$ with zeros, so $B^QA^Q$ is zero at the beginning of fine-tuning. The computation of $W_g^K$ and $W_g^V$ is analogous to $W_g^Q$.

\section{Dataset and Experimental Design}

\revised{To confirm the effectiveness of \model, we evaluate it against baseline models on our file-level code summarization dataset \dataset. In this section, we will first introduce the construction of the \dataset and then the details about the experimental setup, including the evaluation metrics, baseline models, and implementation details.}

\subsection{FILE-CS Dataset}\label{sec:data_collect}
To the best of our knowledge, there is no available dataset for file-level code summarization. Therefore, we construct \dataset, the first dataset for file-level code summarization, containing 98,236 pairs of code files and summaries. We focus on Python in this work and plan to support more programming languages in the future. 
We first collect original data from the github-code\footnote{\url{https://huggingface.co/datasets/codeparrot/github-code}} publicly released in HuggingFace, which consists of a large number of code files from GitHub, totaling 1TB of data. 
Then, we use the following rules to construct our dataset:
\begin{itemize}
    \item We first extract summary information of the file based on manually constructed rules. For the Python language, developers often write comments summarizing the code file in the first paragraph using triple quotation marks, such as \textit{"""}. We extract the content between these triple quotes as the summary for the corresponding files. An example is \texttt{""" Fine-tuning the library models for language modeling on a text file. """}.
    
    \item To ensure the quality of our dataset, we remove files with summaries that are too short or too long. A high-quality summary is considered a concise sentence that accurately describes the function implemented by a file. Therefore, we remove code files with summaries that have fewer than 5 tokens or more than 128 tokens.
    
    \item  To improve the quality of our dataset, we filter samples whose summaries contain automatically generated license/copyright headers. 
    
    \item Finally, we follow previous work \cite{allamanis2019adverse} to deduplicate examples with high similarity and obtain 98,236 <code file, summary> pairs. We split them into 78,588/9,824/9,824 examples for training/development/testing. 

\end{itemize}

We list the statistics about the dataset in Table \ref{data-sta}. The average length of code tokens in file-level code summarization is 1,295, and pre-trained models with a maximum input length of 512 cannot handle the entire sequence for most code files.
\begin{table}[ht]
\caption{Data statistics about \dataset.}
    \centering
    \small
    \begin{tabular}{lcccc}
    \toprule
	Length &  Avg & 25\% & 50\% & 75\% \\
    \midrule
    Source Code &1295.0&564& 890 &1563\\
    Summary &11.0&6&8&12 \\
    \bottomrule
\end{tabular}

\label{data-sta}
\vspace{-0.2in}
\saveSpaceSec
\end{table}

\subsection{Evaluation Metrics}
\subsubsection{BLEU}
BLEU measures the average n-gram precision between the reference sentences and generated sentences, with a brevity penalty for short sentences. The formula to compute BLEU-1/2/3/4 is:
\begin{equation}
\operatorname{BLEU-N}= BP \cdot \exp \sum_{n=1}^{N} \omega_{n} \log p_{n},
\end{equation}
where  $p_n$ (n-gram precision) is the fraction of n-grams in the generated sentences that are present in the reference sentences, and $\omega_{n}$ is the uniform weight 1/N.
Since the generated summary is very short, high-order n-grams may not overlap. We use the +1 smoothing function~\cite{LinO04}. BP is a brevity penalty given as:
\begin{equation}
    BP=\left\{\begin{array}{cl}
                    1 & \text { if } c>r \\
                    e^{(1-r / c)} & \text { if } c \leq r
            \end{array}\right.
\end{equation}
Here, c is the length of the generated summary, and r is the length of the reference sentence.

\subsubsection{Rouge-L}

Based on the longest common subsequence (LCS), Rouge-L is widely used in text summarization. Instead of using only recall, it uses the F-score which is the harmonic mean of precision and recall values. Suppose $A$ and $B$ are generated and reference summaries of lengths c and r respectively, we have:
\begin{equation}
    \left\{\begin{array}{cl}
        P_{Rouge-L}=\frac{L C S(A, B)}{c}\\
        R_{Rouge-L}=\frac{L C S(A, B)}{r}\\
     \end{array}\right.
\end{equation}

$F_{Rouge-L}$, which indicates the value of $\operatorname{Rouge-L}$, is calculated as the weighted harmonic mean of $P_{Rouge-L}$ and $R_{Rouge-L}$:
\begin{equation}\label{equ:rouge}
        F_{Rouge-L}=\frac{\left(1+\beta^{2}\right) P_{Rouge-L} \cdot R_{Rouge-L}}{R_{Rouge-L}+\beta^{2} P_{Rouge-L}}
\end{equation}

\subsubsection{Meteor}
Meteor is a recall-oriented metric that measures how well the model captures the content from the references in the generated sentences and has a better correlation with human judgment. Suppose $m$ is the number of mapped unigrams between the reference and generated sentence with lengths $c$ and $r$ respectively.
Then, precision, recall, and F are given as:
\begin{equation}\label{equ:meteor_1}
        P=\frac{m}{c},\,\,
        R=\frac{m}{r},\,\,
        F=\frac{P R}{\alpha P+ (1-\alpha)R}
\end{equation}
The sequence of mapping unigrams between the two sentences is divided into the fewest possible number of ``chunks''. This way, the matching unigrams in each ``chunk'' are adjacent (in two sentences) and the word order is the same. The penalty is then computed as:
\begin{equation}\label{equ:meteor_2}
\text {Pen}=\gamma \cdot \text { frag }^{\beta}
\end{equation}

\noindent where $\text {frag}$ is a fragmentation fraction:
$\text{frag}=ch/m$, where $ch$ is the number of matching chunks and $m$ is
the total number of matches.

\subsection{Baseline Models}
We compare \model with several publicly released state-of-the-art pre-trained code models, including full self-attention and other sparse Transformer models. For full self-attention Transformer models,we consider CodeBERT~\cite{FengGTDFGS0LJZ20}, GraphCodeBERT~\cite{guo2021graphcodebert}, PLBART~\cite{Ahmad2021UnifiedPF}, CodeT5~\cite{Wang2021CodeT5IU}, and UniXcoder~\cite{guo-etal-2022-unixcoder}. And for other sparse Transformer models, we consider LongFormer~\cite{beltagy2020longformer}, BigBird~\cite{zaheer2020big}, SASA~\cite{liu2022understanding}, LongT5~\cite{guo-etal-2022-longt5}, and LSG Attention~\cite{condevaux2023lsg}. \revised{Specifically, for a fair comparison, we evaluated both the standard LongT5 and LongT5 initialized from UniXcoder. Besides, LSG Attention is initialized from UniXcoder.} Additional details about these baselines can be found in Section \ref{sec:related_work}.

\subsection{Implementation Details} 
We set the maximum length of the code sequence to 512 and 4096 for non-sparse and sparse models, respectively. Additionally, we set the maximum length of summaries to 128 and use the Adam optimizer 
with batch size 16 and learning rate $5e^{-5}$. 
For the sparse models, we adopt the parameters of  UniXcoder~\cite{guo-etal-2022-unixcoder} to initialize the models. For \model, we set the window size $w$, the maximum size of global tokens $|G|$ identifier tokens $|I|$, and rank $k$ in low-rank adaptation as 128, 64, 768, and 8, respectively. During fine-tuning, we set the training epochs as 10 and perform early stopping on the development set. During calculating the Rouge-L and Meteor scores, $\beta$ in equation~\ref{equ:rouge} is set to $1.2$ as in existing studies \cite{zhangretrieval20,WanZYXY0Y18} and the default values of $\alpha$ in equation~\ref{equ:meteor_1} and $\beta, \gamma$ in equation~\ref{equ:meteor_2}
are 0.9, 3.0, and 0.5, respectively.

\section{Results}
\revised{We compare the \model with other baselines on the \dataset to evaluate the performance of the SparseCoder and design a series of ablation experiments to analyze the impact of different components of the SparseCoder and methods adopted in the SparseCoder on the overall performance. To explore the efficiency of the model, We analyze memory consumption and performance. In addition, we conduct human evaluation and case studies. In this section, We will present the experimental results in detail and give an in-depth analysis.}

\subsection{Comparison with Baselines}
To evaluate the effectiveness of the \model, we conduct a series of experiments by comparing it with five state-of-the-art pre-trained code models and six sparse transformer-based models. We present the results in Table~\ref{table:main-result}. The first group of results in the table shows the performance of non-sparse models and the second shows the performance of sparse models.
\begin{table}[ht]
\caption{Performance of different models on FILE-CS.}
\small
    \centering
    \begin{tabular}{lccc}
    \toprule
	Model & BLEU & Rouge-L & Meteor \\
        \midrule
        GraphCodeBERT &16.9&26.2&21.8 \\
        CodeBERT &17.4&26.7&22.2 \\
        PLBART &18.6&28.0&25.2 \\ 
        CodeT5 &19.1&28.7&25.2 \\
        UniXcoder &19.3&29.3&25.0 \\
        \midrule
        BigBird &19.7&29.3&25.2 \\
        SASA &19.9&29.7&25.4 \\
        LongFormer &20.1&30.4&26.1 \\
        LongT5 &18.6&25.3&24.3 \\
        Unix-LongT5$^{\mathrm{*}}$ &20.6&29.3&26.3 \\
        
        LSG Attention &20.8&30.5&26.6 \\

        \midrule
        \model &\bf{21.4}&\bf{32.2}&\bf{27.6} \\
        
    \bottomrule
    \multicolumn{4}{l}{$^{\mathrm{*}}$LongT5 initialized from UniXcoder.}
\end{tabular}
\label{table:main-result}
\vspace{-0.1in}
\saveSpaceSec
\end{table}

The results in Table \ref{table:main-result} demonstrate that sparse Transformer models perform better than models using a full self-attention mechanism. The main reason is that non-sparse models have to truncate the input length due to memory constraints, resulting in the loss of important information. It also indicates that file-level code summarization requires models supporting long sequence inputs. After leveraging the code structure to a sparse attention matrix, \model outperforms other sparse models and achieves
4.3\%$\sim$19.1\% improvement in overall score, which demonstrates the effectiveness of our proposed sparse attention patterns. We also give two case studies in Section \ref{sec:casestudy} for qualitative analysis. 
In summary, results show that the \model can capture complex code structures and therefore generate correct summaries. 
\saveSpaceSec

\subsection{Component Analysis}\label{section:ablation}
We conduct additional experiments to analyze the impact of different components of \model and methods adopted in \model on overall effectiveness and efficiency.  The results are shown in Table \ref{table:ablation}.

\begin{table}[ht]
\caption{Result of ablation study.}
    \small
    \centering
    \setlength{\tabcolsep}{1.4mm}{
    \begin{tabular}{lcccc}
    \toprule
		Model  & Parameters & BLEU & Rouge-L & Meteor \\
        \midrule
        \model &129.9M&\bf{21.4}&\bf{32.2}&\bf{27.6} \\
        \ - w/o global attention &129.9M&20.4&30.8&26.6 \\
        \ - w/o identifier attention &129.9M&20.8&31.2&27.1 \\
        \ - replace random tokens &129.9M&20.5&31.0&26.7 \\
        \ - w/o low-rank	&129.5M&20.9&31.5&26.8\\
         \ - w/ new-proj &150.7M&21.5&32.2&27.5 \\

    \bottomrule
    \end{tabular}
    }
\label{table:ablation}
\vspace{-0.1in}
\saveSpaceSec
\end{table}

First, to understand the impact of the different sparse attention patterns on the overall performance, we conduct additional experiments by removing or replacing various patterns with random tokens. From Table~\ref{table:ablation}, we can see that the average score of three metrics drops from 27.1\% to 26.0\% and 26.4\% when removing global attention (\textbf{w/o global attention}) and identifier attention (\textbf{w/o identifier attention}), respectively, which reveals the importance of these attention patterns. \newrevised{Notably, we set the maximum size of global tokens
and identifier tokens as 64 and 768, respectively. Therefore, the parameters with or without global attention remain almost unchanged.} These results indicate that global and identifier attention patterns contribute to the effectiveness of \model. Furthermore, we find that replacing global tokens and identifier tokens with random tokens (\textbf{replace random tokens}) results in worse performance than the original \model, indicating that the benefit of the identifier attention pattern is not solely due to increased attention calculation. 

\revised{Then, we conduct additional experiments to study the impact of different methods adopted in \model on overall effectiveness and efficiency. From the results shown in the last two rows of Table \ref{table:ablation}, we can see that: (1) Sharing the parameters across all attentions (\textbf{w/o low-rank}) can reduce the model size but results in a decrease in performance. 
 (2) Using two distinct sets of projections Q, K, and V, for global attention and other attentions (\textbf{w/ new-proj}) 
brings certain performance improvements, but also introduces many additional parameters, approximately 16\% of the original model. It highlights the effectiveness of utilizing LoRA in global attention. It illustrates that the combination of various attention patterns and the LoRA method in our \model model can reduce the number of parameters while improving performance.}

\revised{In summary, the experimental results of the ablation study demonstrate the effectiveness of our proposed global and identifier attention mechanism and the improvement of efficiency by adopting the LoRA method.} 

\subsection{Memory and Performance Analysis}
\label{sec:memory}
\begin{figure}[ht]
    \centering
    \includegraphics[width=0.9\linewidth]{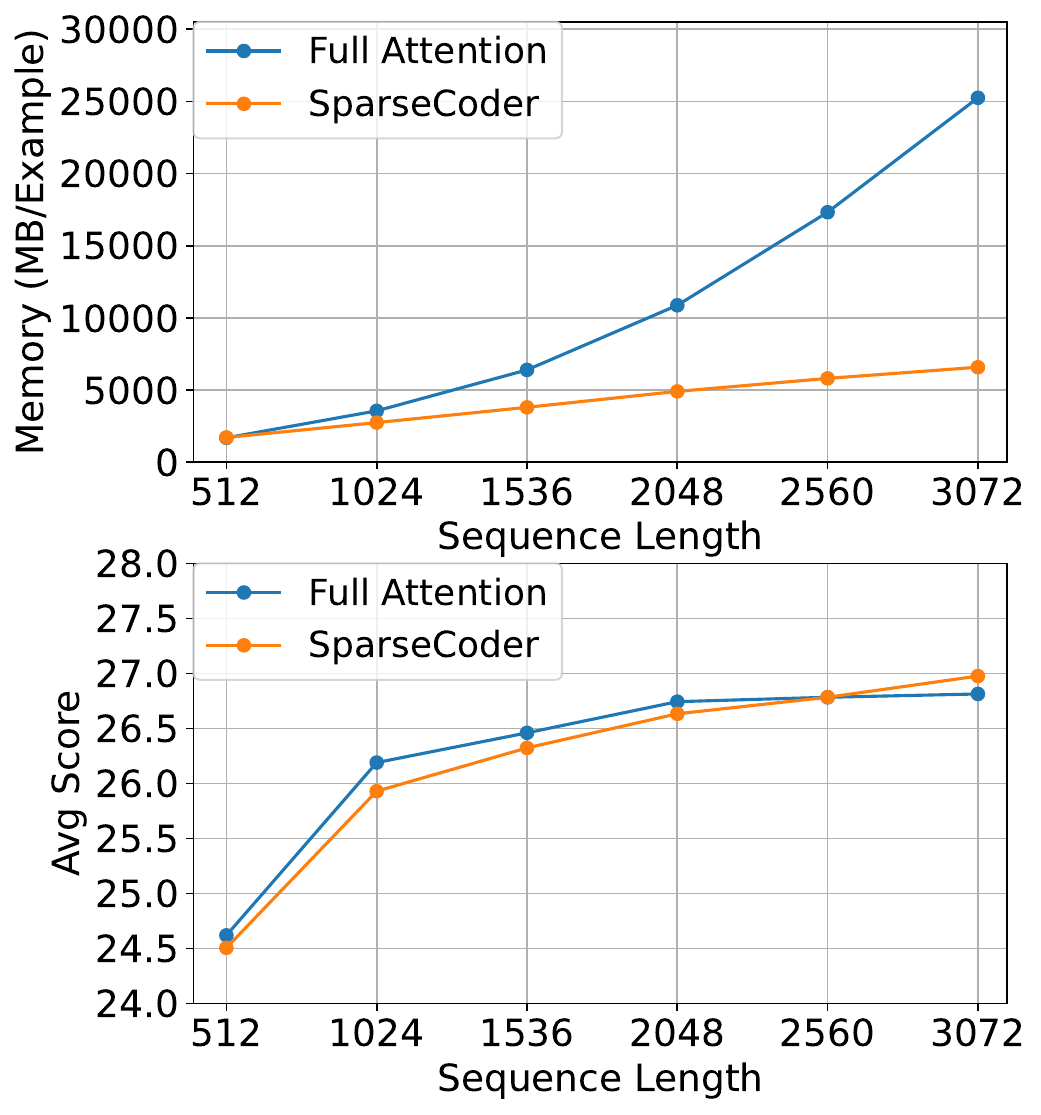}
    \saveSpaceSec
    \caption{The effect of input sequence length on memory usage and the average score of the three metrics. }
\label{fig:plot}
\vspace{-0.1in}
\end{figure}

 To investigate the memory cost and performance under different input sequence lengths, we conduct a series of experiments comparing \model with the full self-attention model UniXcoder~\cite{guo-etal-2022-unixcoder} released by Guo et al. on \dataset. We present the experimental results in Figure~\ref{fig:plot}. Our analysis reveals that as the sequence length increases, the memory cost of the full self-attention model increases at a faster rate (quadratically) than \model (linearly). 
However, their performance under varying sequence lengths is comparable. Although full self-attention slightly outperforms the \model, its memory cost is very high. An exception is when the sequence length is very long ($\ge 
 2560$), \model performs better than full self-attention models. 
The main reason may be that full self-attention can lead to redundant or noisy information in long code sequences. However, the \model can capture semantic information of code more accurately and better understand it by leveraging designed sparse attention patterns.

\subsection{Human Evaluation}
\revised{Besides textual similarity-based metrics, we
conduct a human evaluation to evaluate the semantic similarity of the summaries generated by \model and other baselines, following the previous work~\cite{hu2020deep,iyer2016summarizing,leclair2020improved,liu2019automatic,shi2021cast}. 
Specifically, we select the best-performing models UniXcoder among the full self-attention models and LSG Attention among the other sparse models for human evaluation.}

\revised{We first randomly choose 100 Python files from the testing sets of \dataset and their summaries generated by three models. In particular, we invite four volunteers with more than three years of software development experience and excellent English ability. Then each volunteer is asked to assign scores from 0 to 4 (the higher, the better) to the generated summary from three aspects: similarity of the generated summary and the ground truth summary, naturalness (grammaticality and fluency, ignoring its correctness), and informativeness (the amount of content carried over from the input source code to the generated summary, ignoring fluency). Each summary is evaluated by four volunteers, and the final score is the average of them.}

\revised{The results shown in Table \ref{table:human-evaluation} demonstrate that \model outperforms other baseline models in all three aspects, which means that
\model tends to generate correct and natural summaries with comprehensive semantics, further 
validating the effectiveness of our \model model.}

\begin{table}[ht]
\caption{Results of human evaluation (standard deviation in parentheses). }
\small
    \centering
    \begin{tabular}{lccc}
    \toprule
	Model &Informativeness&Naturalness&Similarity\\
        \midrule
        SparseCoder	&\textbf{2.085}(0.98)&\textbf{3.175}(0.60)&\textbf{2.353}(1.19) \\
        LSG Attention	&2.065(1.02)&3.094(0.70)&2.160(0.93) \\
        UniXcoder	&2.020(1.00)&3.075(0.70)&2.133(0.93) \\

    \bottomrule
\end{tabular}

\saveSpaceSec

\label{table:human-evaluation}
\vspace{-0.2in}
\end{table}

\subsection{Case Studies}\label{sec:casestudy}
\begin{figure*}[ht]
    \centering
    \includegraphics[width=0.99\linewidth]{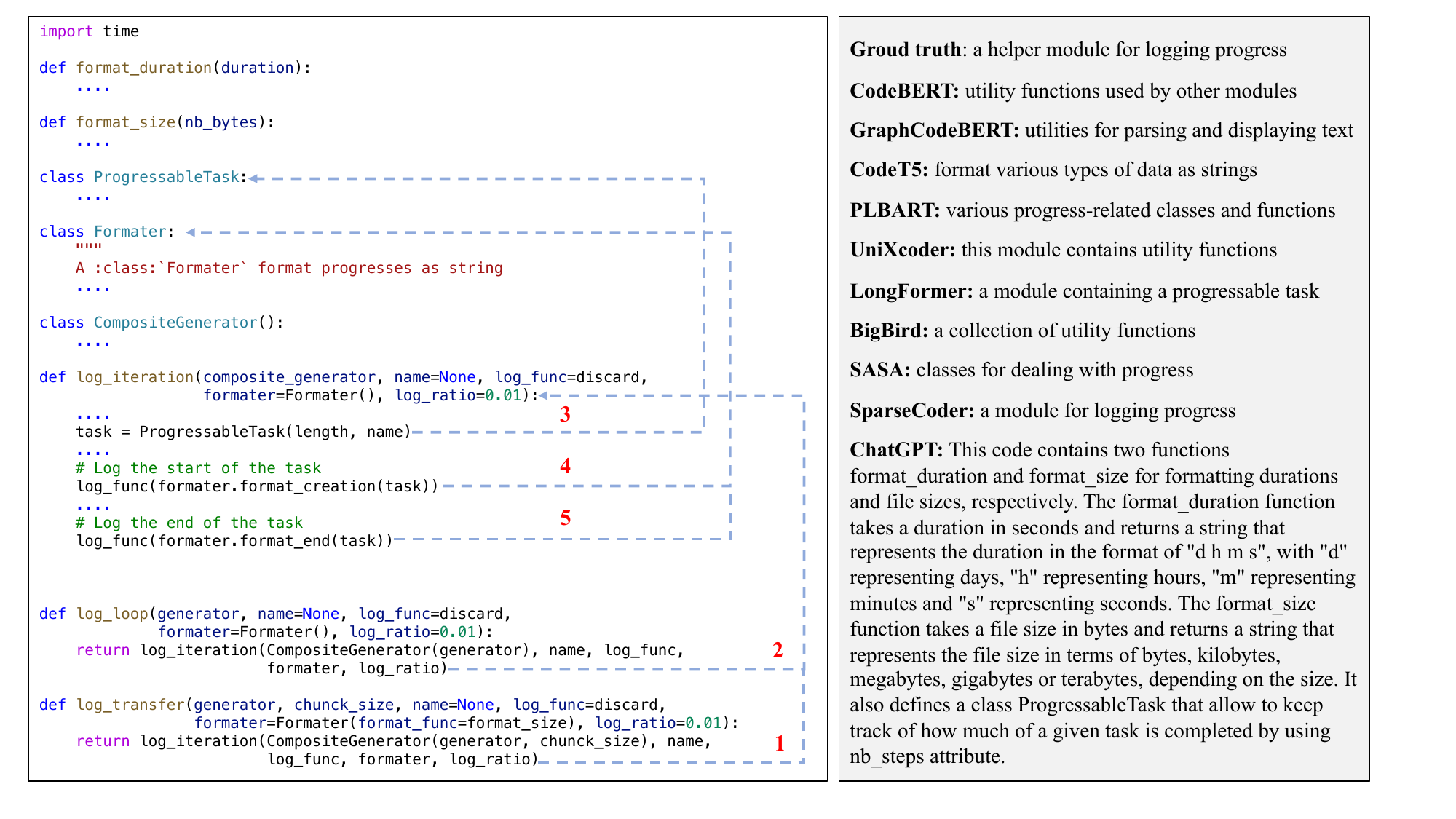}
    \caption{An example on \dataset dataset and the predictions from different models. The input code file is displayed on the left, and the predictions from the models are provided on the right. Arrows indicate the key call relationship.}
    \label{fig:case}
    \vspace{-0.1in}
\end{figure*}
\begin{figure*}[t]
    \centering
    \includegraphics[width=0.95\linewidth]{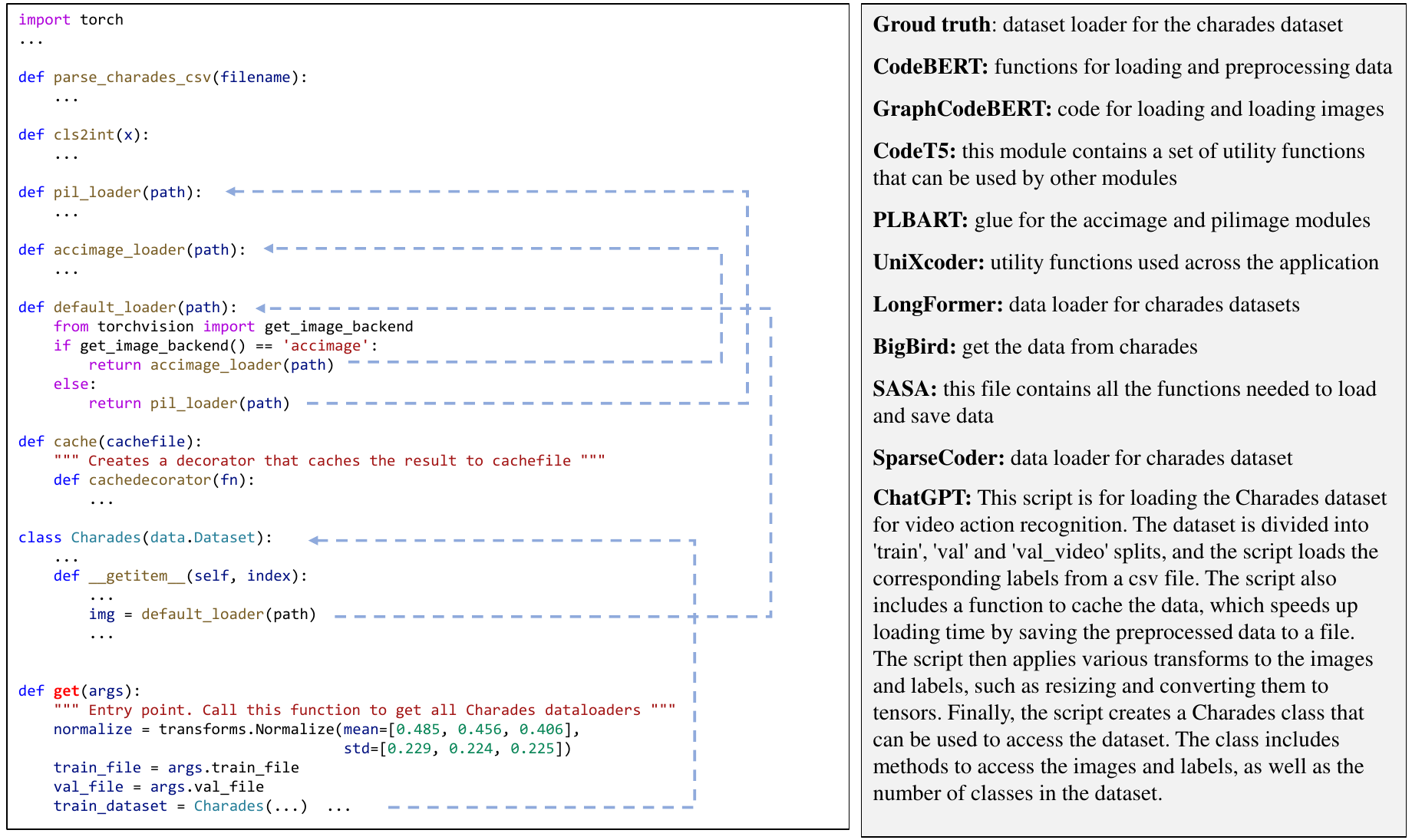}
    \caption{Another example on \dataset dataset and the predictions from different models. The input code file is displayed on the left, and the predictions from the models are provided on the right. Arrows indicate the key call relationship.}
    \label{fig:case2}
    \vspace{-0.15in}
\end{figure*}

\subsubsection{Case Study 1}
Figure \ref{fig:case} shows an example of the generated summaries from different models. It is observed that the \model generates a correct summary compared to other models.
(1) As full self-attention models are constrained by input length, they are unable to detect and process logging-related functions located beyond the truncation point. As a result, they are incapable of generating file-level summaries related to logging.
(2) Despite the presence of numerous logging-related functions at the end of the file, the complex calling relationships between them pose a challenge for other sparse models. To correctly generate a summary, the model must first identify the primary function of the file as \texttt{log\_iteration} (key clues are marked in 1-2), and then understand that the information recorded in this function is related to the \texttt{task} variable, which is an instance of \texttt{ProgressableTask} (key clues are marked in 3-5). The \model is able to handle these identifier relationships and produce accurate results effectively.
(3) We also attempted using ChatGPT\footnote{\url{https://chat.openai.com/}} with the prompt ``Please generate a file-level summary of the following code'' and found that while ChatGPT tends to summarize each function and class of the file well, it struggles to effectively understand the relationships between them and generate a concise file-level summary. 
Additionally, when the code input is too long, ChatGPT returns ``The message you submitted was too long''. It highlights the importance of efficient processing of long code sequences.

\subsubsection{Case Study 2}\label{sec:example2}
We show the second case study in Figure~\ref{fig:case2}. It can be observed that the \model generates a correct summary compared to other models. The crucial aspect is that the model must examine the last function named \texttt{get} in the file, which is in red font, to acquire complete information about the code. In addition, the model needs to infer the main functionality by the calling relationships within the code. Through utilizing the three attention patterns, \model is able to capture these relationships and generate the most effective summary. 

\saveSpaceSec
\section{Generality of SparseCoder}\label{sec:generality}
\revised{While the paper mainly studies the task of file-level code summarization, \model can be used for not only this task. It can be applied to other code understanding tasks. Furthermore, it can also be evaluated on different datasets. In this section, we will introduce experiments on the generality of \model, including its performance on other tasks datasets.}

\begin{table}[ht]
\caption{Performance of different models on Code Clone Detection and Code Search. }
\small
    \centering
    \begin{tabular}{lccc}
    \toprule
	Model & \# Parameters&BCB & CSN-Py\\
        \midrule
        CodeBERT &125.8M&82.5&80.8 \\
        GraphCodeBERT &125.8M&83.3&82.3 \\
        UniXcoder &127.1M&86.3&84.1 \\
        BigBird &128.6M&89.4&84.5 \\
        SASA &128.6M&90.1&84.5 \\
        LongFormer &150.7M&91.4&84.7 \\
        \midrule
        \model &129.9M&\bf{92.8}&\bf{85.5} \\
        \ - w/\ \ \  new-proj &150.7M&92.7&85.4 \\
        \ - w/o low-rank &129.5M&92.2&84.8 \\
    \bottomrule
    \vspace{-0.15in}
    \end{tabular}

\label{table:other-result}

\saveSpaceSec
\end{table}

\begin{table}[ht]
\caption{Performance of different models on Filtered-CSN.}
\small
    \centering
    \setlength{\tabcolsep}{1.3mm}{
    \begin{tabular}{lccccccc}
    \toprule
	Model &Java&Pytho&Go&Php&JavaScript&Ruby&Overall \\
        \midrule
        UniXcoder &13.8	&14.9&11.2&19.9	&13.7&12.5&14.3 \\
        LongFormer &14.5&15.0&11.4&20.3&14.0&13.7&14.8 \\
        
        \model &\bf{14.9}&\bf{15.4}&\bf{11.8}&\bf{20.6}&\bf{14.3}&\bf{14.6}&\bf{15.3} \\
        
    \bottomrule
    \vspace{-0.15in}
    \end{tabular}
    }
\label{table:result-on-CSN}

\saveSpaceSec

\end{table}
\subsection{Generalize to Other Tasks}
We conduct additional experiments to evaluate its performance on code clone detection and code search tasks. To assess the efficacy of our model, we employ the BigCloneBench \cite{svajlenko2014towards} (BCB) dataset for code clone detection and CodeSearchNet~\cite{guo2021graphcodebert} in Python (CSN-Py) for code search.
Specifically, to evaluate the efficiency of modeling long code sequences, we filter out code shorter than 512 tokens. 
The details about the datasets are introduced as follows.

\subsubsection{BCB}
BigCloneBench consists of a large collection of open-source Java projects from GitHub, which are carefully analyzed to identify code clones and code similarity instances. The dataset consists of 6 million true clone pairs and 260 thousand false clone pairs in total.
After the filtering step as described above, the training, validation, and testing sets contain 28,595, 15,753, and 10,296 samples, respectively.

\subsubsection{CSN-Py}
The CodeSearchNet dataset is a large-scale dataset that includes code snippets and the corresponding documentation from multiple programming languages. We extract from the Python data in CodeSearchNet and filter the data as described above. The training, validation, and testing sets contain 9,477, 593, and 627 samples, respectively. And the code candidates pool contains 1,980 code snippets.

We use the F1 score and MRR as evaluation metrics for code clone detection and code search, respectively, and present the results in Table~\ref{table:other-result}. The results illustrate that SparseCoder has the best performance in both two tasks, showing that our proposed identifier-aware sparse Transformer model can obtain better code representations, leading to excellent performance when applied to various downstream tasks. 
Additionally, we perform ablation studies similar to the experiments conducted in Section~\ref{section:ablation}.
The results are consistent with the previous results in Table~\ref{table:ablation}, showing that the combination of various attention patterns and the adoption of the low-rank method can effectively reduce model size while simultaneously improving performance across various code understanding tasks, further proving the effectiveness and generalizability of our \model model.

\subsection{Generalize to Other Datasets}
\revised{We conduct further experiments to evaluate its performance on the widely-used CodeSearchNet (CSN) dataset, which comprises multiple programming languages.
Applying the filtering rules described in section \ref{sec:data_collect}, we obtained 25,981 examples for training, 1,228 for validation, and 1,534 for testing. Specifically, we select UniXcoder and LongFormer, which have outstanding performance in the full self-attention and sparse models, and then compare them with \model on the filtered CodeSearchNet dataset.
We choose BLEU as the evaluation metric, and the results are presented in Table~\ref{table:result-on-CSN}. The results show that our \model model outperforms other baseline models across a variety of programming languages, which strongly proves the effectiveness of the \model. } 

\section{Threats to Validity}
\revised{One threat to validity is the current limitation of our dataset, which includes only the mainstream programming language, Python. However, the construction approach of our \dataset dataset is not language-specific, and it can be adapted for other programming languages. In future research, we plan to extend the \dataset dataset into a multilingual dataset, enhancing its reliability and practicality.}
\revised{Another potential threat is that our method has been primarily applied to small models with a limited number of parameters (e.g., UniXcoder with just over 100 million parameters). Currently, various large language models with billions of parameters have been proposed. In future research, we intend to explore the performance of our method on these large-scale models to evaluate its scalability and applicability in this context. }

\section{Conclusion}
\incSpace
To support automatic file-level code summarization, we present \model, an identifier-aware sparse Transformer for efficiently modeling long code sequences. \model leverages code structures to improve the self-attention mechanism by introducing three types of sparse attention patterns: local, global, and identifier attention. Experiments show that the \model significantly outperforms previous work including both full self-attention and other sparse models. The ablation studies show that the combination of various attention patterns and the adoption of the LoRA method in the \model is effective. Further experiments also show that the \model can be generalized to other tasks such as code clone detection and code search. 

\printbibliography
\balance
\end{document}